\documentstyle[sprocl]{article}

\def\beq{\begin{equation}} 
\def\eeq{\end{equation}}
\def\beqa{\begin{eqnarray}} 
\def\eeqa{\end{eqnarray}} 
\def\lla{\left\langle}
\def\rra{\right\rangle}

\begin{document}

%% net cover
%\addtolength{\baselineskip}{.2cm}

\begin{flushright}
UR-1537\\
ER/40685/919\\
July 1998
\end{flushright}

\vspace*{.2in}

\begin{center}
{\bf Gauge Mediated Supersymmetry Breaking
 and Spontaneous CP Violation as a Solution to the Strong CP Problem}$^\star$

\vspace*{.4in}
{\bf Otto C.W. Kong }

\vspace*{.4in}
{\it Department of Physics and Astronomy,\\
University of Rochester, Rochester NY 14627-0171.}

\vspace*{.8in}
{Abstract}
\end{center}

\noindent
The origin of CP violation is a major mystery, especially in 
relation to the strong CP problem. CP being a spontaneously 
broken symmetry could provide an elegant solution. However, such models
have difficulty making themselves compatible with low-energy
supersymmetry, which is popularly accepted as the solution to the hierarchy
problem. We demonstrate that a certain class of low scale supersymmetric 
``Nelson-Barr'' type models can solve the strong and supersymmetric CP 
problems while at the same time generating sufficient weak CP violation 
in the $K^{0}-\bar{K}^{0}$ system. Gauge-mediated supersymmetry 
breaking is used to provide the needed squark mass degeneracies and 
$A$-term proportionalities; though that proves to be still insufficient for a 
generic Nelson-Barr model. The workable model we consider here,
essentially a supersymmetric version of the aspon model, has
the Nelson-Barr mass texture enforced by a U(1) gauge symmetry, 
broken at the TeV scale. The resulting model is predictive with rich 
phenomenology soon to be available. Feasibility of the model considered 
is established by a detailed renormalization group studies.

\vfill
\noindent --------------- \\
$^\star$ Talk given at  PASCOS-98, Northeastern University, March 1998, and  at the 4th International Workshop on Particle Phenomenology, National Sun Yat-Sen University, Taiwan 
June 1998 --- submission for proceedings.  

\clearpage

%%%%%%%%%%%%%%%%%%%%%%%%%%%%%%%%%%%%

%\addtolength{\baselineskip}{-.2cm}
\addtocounter{page}{-1}

\title{ Gauge Mediated Supersymmetry Breaking
 and Spontaneous CP Violation as a Solution to the Strong CP Problem }

\author{ Otto C.W. Kong}

\address{Department of Physics and Astronomy,\\
University of Rochester, Rochester NY 14627-0171.}

\maketitle 
\abstracts{
The origin of CP violation is a major mystery, especially in 
relation to the strong CP problem. CP being a spontaneously 
broken symmetry could provide an elegant solution. However, such models
have difficulty making themselves compatible with low-energy
supersymmetry, which is popularly accepted as the solution to the hierarchy
problem. We demonstrate that a certain class of low scale supersymmetric 
``Nelson-Barr'' type models can solve the strong and supersymmetric CP 
problems while at the same time generating sufficient weak CP violation 
in the $K^{0}-\bar{K}^{0}$ system. Gauge-mediated supersymmetry 
breaking is used to provide the needed squark mass degeneracies and 
$A$-term proportionalities; though that proves to be still insufficient for a 
generic Nelson-Barr model. The workable model we consider here,
essentially a supersymmetric version of the aspon model, has
the Nelson-Barr mass texture enforced by a U(1) gauge symmetry, 
broken at the TeV scale. The resulting model is predictive with rich 
phenomenology soon to be available. Feasibility of the model considered 
is established by a detailed renormalization group studies.
 }

The sources of CP violation is still a major puzzle. While the only
experimentally observed CP violating effect, in the $K^{0}-\bar{K}^{0}$ 
system, is compatible with the Kabayashi-Maskawa (KM) prescription
within the standard model, there are evidents for a more elaborate 
theory of CP violation. Among them, the strong CP problem is the most
pressing. A potential strong interaction contribution to the neutron electric 
dipole moment and leads to the very stringent experimental contraint
\begin{equation}
\bar{\theta} <   10^{-9}~.
\end{equation}
Here, $\bar{\theta}$ is  the physical combination of the strong CP 
parameter $\theta$ and the CP phase(s) of the color fermion mass matrices.
In the case of the supersymmetric standard model, this is to be given by
\begin{equation}
\bar{\theta} = \theta - {\mathrm arg\, det} M_q - 3\, {\mathrm arg} M_g~,
\label{thetabarsusy}
\end{equation} 
where $ M_q$ is the quark mass matrix and $M_g$ the gluino mass.
The  incredible smallness of this {\it a priori} arbitary parameter demands
an explanation.
With supersymmetry (SUSY), there are also a large number of admissible
CP violating phases among the soft SUSY breaking parameters (SSBP),
which are also constrained to be quite small ($\sim 10 ^{-3}$).

Using spontaneous CP violation to solve the strong CP problem
was pioneered by Nelson and Barr\cite{NB}. Extending the approach
to a  SUSY model could take care of the SUSY CP phases easily.
Attempts to realize the Nelson-Barr mechanism in SUSY
models\cite{BMS} have, however, run up against a formidable
difficulty: There generically exist potentially large 1-loop contributions 
to $\bar{\theta}$ in these models. As discussed at length in 
Ref.~\cite{DKL}, one requires an exceptionally high degree
of proportionality of the soft SUSY breaking trilinear scalar couplings
to their associated Yukawa couplings as well as degeneracy among
the soft squark mass terms for each charge and color sector, if these 
contributions are to be sufficiently suppressed. The degree
of proportionality and degeneracy required among the SSBP
is very difficult to maintain due to the effects of renormalization.

A particularly interesting modified version of the Nelson-Barr type model
is given by the aspon model\cite{aspon}. Here the extra symmetry 
needed to enforce the Nelson-Barr mass texture is promoted to a new
$U(1)$ gauge symmetry. This $U(1)_A$ gauge boson (aspon) then
mediates a tree level CP violating $K^{0}-\bar{K}^{0}$ mixing. This 
allows the effective KM phase to be small. A preliminary study suggested
a SUSY version of the model has a much better chance to provide a
feasible solution\cite{pp}. The detail model is constructed in the
framework of gauge mediated SUSY breaking (GMSB), and demonstrated 
to work, through a careful renormalization group (RG) study\cite{p}.

The model have CP spontaneously broken at low energies (of order a 
TeV). It has an extra right-handed down quark superfield $\bar{D}$ 
together with its mirror $D$ coupling to the ordinary down quarks via the 
superpotential
\begin{equation}
W_d = Y_d^{ij}Q_j\bar{d}_i H_d + \mu_{{\scriptscriptstyle D}} D\bar{D} +
\gamma^{ia}D\chi_a \bar{d}_i~,\label{Wd}
\end{equation}
where the VEVs of the scalar components of the two $\chi_a$'s contain a 
relative phase, thus breaking CP. $D$ and $\chi_a$ have opposite
$U(1)_A$ charges. The down-sector quark mass matrix can be written
as
\begin{equation}
m_q = \left( \begin{array}{cc}
m_d & x  \mu_{{\scriptscriptstyle D}} \mbox{\boldmath $a$} \\
0 & \mu_{{\scriptscriptstyle D}}
\end{array}\right)~,\label{mf}
\end{equation}
and squark mass matrices as
\beqa
 \widetilde{M}^2_{RL} & = &
 \left( \begin{array}{cc}\bar{A}_d m_q(1 + \mu \tan\beta)  
 & \bar{A}_{\gamma} x \mu_{{\scriptscriptstyle D}}  \mbox{\boldmath $a$}\\
0 & \bar{A}_{\gamma} \mu_{{\scriptscriptstyle D}}\end{array} \right) 
+   \left( \begin{array}{cc} \left\langle H_d \right\rangle \delta\! A_d & 
 x \mu_{{\scriptscriptstyle D}} \delta\! A_{\gamma}  \mbox{\boldmath $c$}\ 
 -\gamma^{ia} \lla F_{\chi_a}\rra \\
 0 & \mu_{{\scriptscriptstyle D}} (B_{{\scriptscriptstyle D}}- \bar{A}_{\gamma})
 \end{array} \right) \; , \\
 \widetilde{M}^2_{RR} & =&  \tilde{m}^2_{R} I
+   \left( \begin{array}{cc}\delta\!  \tilde{m}^2_{R} & 0 \\
0 & \delta\!  \tilde{M}^2_{R}\end{array} \right)    + m_q  m_q^{\dag}
\; , \\
 \widetilde{M}^2_{LL} & =&  \tilde{m}^2_{L} I
+   \left( \begin{array}{cc}\delta\!  \tilde{m}^2_{L} & 0 \\
0 & \delta\!  \tilde{M}^2_{L}\end{array} \right)   + m_q^{\dag} m_q \; , 
\eeqa
where $a^i = \frac{1}{x \mu_{{\scriptscriptstyle D}}}
\gamma^{ia}\langle\chi_a\rangle$, such that
{\boldmath $a^{\dag} a$}$=1$, and
\beq
\delta\!A_{\gamma} c^i = \frac{1}{x \mu_{\scriptscriptstyle D}} h^{ia}_{\gamma} \lla \chi_a \rra
- \bar{A}_{\gamma}  a^i~,  		\label{dagam}
\eeq
with {\boldmath $c^{\dag} c$}$=1$. Hence, the complex phases are
all contained in three-vectors {\boldmath $a$} and {\boldmath $c$},
and the most dangerous $ \lla F_{\chi_a}\rra$'s.

The spontaneous CP violation (SCPV) is obtained from a superpotential of
$ \chi_{a}$'s together with two oppositely charged $\bar{ \chi}_{a}$'s
and a neutral $\aleph$ superfield,
\beq
W_{\chi} =  \bar{\chi}_a \mu_{\chi}^{ab} \chi_{b}
+ \aleph \bar{\chi}_a \lambda^{ab} \chi_b
+ \lambda_{{\scriptscriptstyle \aleph}} \aleph^3
+ \mu_{{\scriptscriptstyle \aleph}} \aleph^2 \label{wchi} \; ,
\eeq
which generally gives a SUSY perserving vacuum with complex VEV's 
breaking CP. The sector is minimal for anomaly cancellation and SCPV.
Without nonzero SSBP for superfields of the sector, the
$ \lla F_{\chi_a}\rra$'s are then vanishing.
A minimal GMSB scenario\cite{Bor} blind to the $U(1)_A$ is used to 
suppress proportionality violations $\delta\! A_d$, 
$ \delta\! A_{\gamma} $, and  
$ (B_{{\scriptscriptstyle D}}$-$\bar{A}_{\gamma})$, degeneracy 
violations, as well as the SSBP for that lifts the zero $ \lla F_{\chi_a}\rra$'s. The latter is then estimates both analytically and diagrammatically, based on 
RG analysis of all SSBP. The 1-loop $\bar{\theta}$ contributions from 
each terms is calculated through the mass insertion method.
The model predicts a measureable neutron EDM, which could be close to 
the present experimental bound for $x >0.01$, with some region of the 
admissible $x$-values (of the non-SUSY version) plausibly rule out by
having  the  $ \lla F_{\chi_a}\rra$'s too large, and also a very rich spectrum 
of new particles around the TeV scale.

\end{document}